\documentclass[pdftex,a4paper,10pt,twocolumn,prl,reprint]{revtex4}

\usepackage[latin1]{inputenc}
\usepackage[T1]{fontenc}
\usepackage{amsmath}
\usepackage{color,graphicx}
\usepackage{xr-hyper} 
\begin{document}
\title{A plasma solenoid driven by an Orbital Angular Momentum laser beam}
\author{R. Nuter$^{1}$, Ph. Korneev$^{2}$, I. Thiele$^{3}$  and V. Tikhonchuk$^{1}$}

\affiliation{
1 Universit\'e Bordeaux,CNRS,CEA,UMR 5107,33405 Talence, France - 
2 National Research Nuclear University 'MEPhI', Moscow, 115409 Russian Federation, Lebedev Physical Institute, Moscow, 
119333, Russian Federation - 3 Chalmers University of Technology, Departement of Physics, SE-41296 G\"oteborg, Sweden}
PACS Numbers: 
\begin{abstract}
A tens of Tesla quasi-static axial magnetic field can be produced in the interaction 
of a short intense laser beam carrying an Orbital Angular Momentum with an 
underdense plasma. Three-dimensional ``Particle In Cell'' simulations and analytical
model demonstrate that orbital angular momentum is transfered from a tightly focused 
radially polarized laser beam to electrons without any dissipative effect. 
A theoretical model describing the balistic interaction of electrons with laser shows 
that particles gain angular velocity during their radial and longitudinal drift 
in the laser field. The agreement between PIC simulations and the 
simplified model identifies routes to increase the intensity of the solenoidal magnetic field 
by controlling the orbital angular momentum and/or the energy of the laser beam. 

\end{abstract}
\maketitle

The generation of a quasi-static magnetic field in the laser-plasma interaction is a subject of many theoretical \cite{berezhiani1997,hertel2006,shvets2002, naseri2010, liseykina2016,lecz2016,korneev2015,korneev2017,ali2010,lecz2016-2} and experimental studies \cite{najmudin2001,deschamps1970,horovitz1997}. Two approaches are considered: one consists in designing the target 
in such way that the interaction with the laser generates controlled azimuthal 
currents \cite{korneev2015,korneev2017}, another one proposes to transfer 
angular momentum from laser to electrons by using circularly polarized laser beam 
\cite{berezhiani1997,hertel2006,shvets2002, naseri2010, liseykina2016,lecz2016,najmudin2001,deschamps1970,horovitz1997} or laser beam with a
structured spatial shape \cite{ali2010,lecz2016-2}. In  
\cite{berezhiani1997,hertel2006,shvets2002, naseri2010, liseykina2016,lecz2016}, the 
authors consider theoretically the magnetization of a medium exposed to a circularly polarized 
Gaussian laser beam. Plasma magnetization originates from the inverse Faraday effect, 
where the spin angular momentum of a laser beam is transfered to the plasma electrons 
due to dissipation processes such as collisions, ionization or radiation friction. This laser to electron angular momentum transfer has been experimentally observed \cite{najmudin2001,deschamps1970,horovitz1997}. Ali {\it et al.} \cite{ali2010}, consider a linearly polarized 
laser beam carrying Orbital Angular Momentum (OAM) \cite{allen1992} and analytically demonstrate 
that such a laser beam transfers its OAM to electron through the inverse bremsstrahlung 
dissipative process. L\'ecz {\it et al.} \cite{lecz2016-2} and Wang {\it et al.} \cite{wang2014}
numerically model the interaction of a screw-shaped laser pulse with an underdense plasma and 
observe laser to electron OAM transfer in the laser wakefield.

In this letter, we demonstrate that a quasi-static axial magnetic field can 
be generated within a purely optical process, without any dissipative effects. 
It is produced in an underdense plasma irradiated by a radially polarized OAM 
laser beam, which is for example, experimentally designed by 
Li {\it et al.} \cite{li2014}. 
The authors demonstrate their capacity to produce such laser beams with an energy 
ratio between radial and azimuthal components attaining 98\%. 
Our three-dimensional (3D) Particle In Cell (PIC) simulations, modeling the laser-plasma 
interaction, clearly show an orbital angular momentum transfer from laser to electron
and the generation of a strong solenoidal magnetic field. A simplified model describing 
the laser-electron dynamics shows that this transfer originates from the joint 
radial and longitudinal electron motion in the laser field. The agreement observed 
between the 3D PIC simulations and the simplified model provides means for controlling
the magnetic field  with laser parameters.

\begin{figure}[h]
\begin{center}
\includegraphics[width=0.99\linewidth]{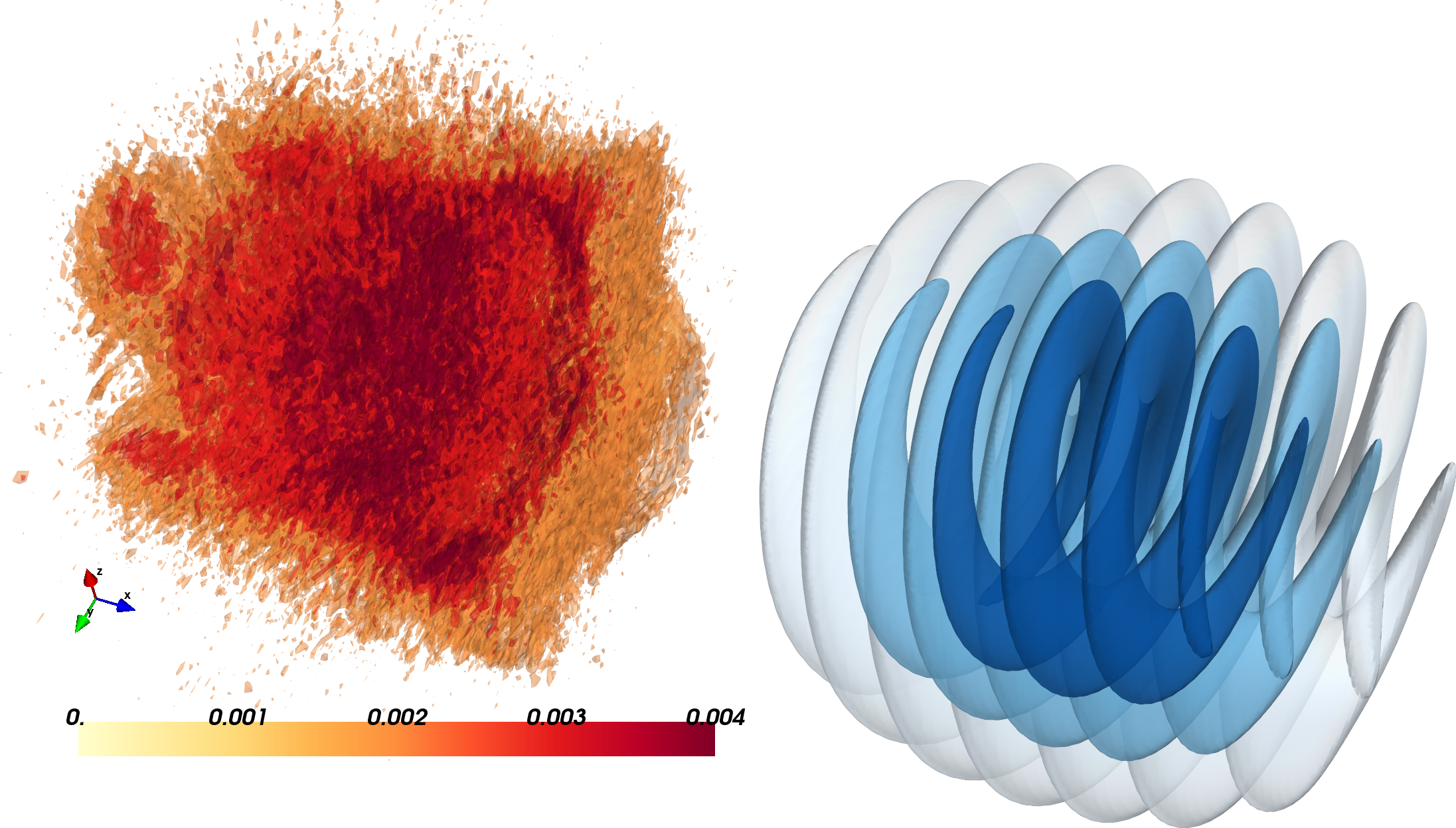}
\caption{(left) Solenoidal magnetic field $B_x$ induced by an OAM $l=1$  laser beam,
 (right) helical energy distribution. The colormap applies to the $B_x$ field.}
\label{fig:field}
\end{center}
\end{figure}

To model a tightly focused radially polarized laser beam carrying OAM, we consider
the numerical algorithm developed by Thiele {\it et al.} \cite {thiele2016}. It consists in 
prescribing the temporal and spatial shape for the electromagnetic fields at the focal 
point $x=x_0$ in vacuum, and then solving the Maxwell's equations in vacuum to 
compute the electric and magnetic fields components 
at the box boundaries ($x$=0 plane). This method provides the electromagnetic fields consistent with the Maxwell's equations 
and is valid beyond the paraxial approximation for laser beam. The radially
polarized OAM laser beam is prescribed at the focal point $x=x_0$ in the cylindrical 
coordinates ($r, \theta, x$):
\begin{equation}
\vec{E}(r,\theta,t, x_0) = E_0 \cdot g(t)\cdot f(r)  \cos{(l\theta - \omega_0 t)} \cdot \vec{e}_{r}
\label{eq:laser}
\end{equation}
with the radial distribution $f(r) = C_l(r/w_0)^{|l|} e^{-(r/w_0)^2}$, the temporal envelop
$g(t)=\cos^2(\pi \frac{t-t_0}{\tau})$ in the time interval $|t-t_0| < \tau/2$, 
the laser OAM $l$, the focal beam waist $w_0 =$ 2 $\mu$m, the pulse duration $\tau$ equal to
6 optical periods (T), 
the central time $t_0$, the laser amplitude $E_0$, the normalization factor 
$C_l = \sqrt{2^{|l|+1}/|l|!}$ and the radial unit vector $\vec{e}_{r}$. 
The laser frequency $\omega_0$ = 2.3 $\times 10^{15}$ $s^{-1}$ corresponds to 
the laser wavelength = 0.8 $\mu$m. The electromagnetic energy 
distribution for the $l=1$ laser beam is illustrated in Fig. \ref{fig:field} (right panel). 
It is characterized by two entangled helices with electric field vectors 
directed along the same radial direction.

Interaction of such a laser beam with a collisionless underdense plasma is studied with 
the 3D PIC code OCEAN \cite{nuter2013}. The numerical box is composed of 640 cells along the 
longitudinal axis $x$, and 512$\times$512 cells in the transverse plane ($y$,$z$), 
with a spatial resolution of 19 nm. Absorbing conditions for the electromagnetic fields 
and particles 
are defined at the box boundaries. The laser pulse with an OAM $l=1$ and an intensity 
2$\times$10$^{18}$ W/cm$^2$, corresponding to the adimensionned field amplitude 
$a_0 = eE_0/m_e\omega_0c=1$ ($m_e$ is the electron mass, $e$ is the electron charge and $c$ is 
the light speed in vacuum), is injected into the numerical box from the left border. The total laser energy 
is 2.5 mJ. The plasma composed of electrons and protons with an initial density equal 
1.74 $\times 10^{19}$ cm$^{-3}$, has a cylindrical spatial shape with a 1.52 $\mu$m length and 
a 6.9 $\mu$m diameter. Ten macro-particles 
per cell are considered.

\begin{figure}[h]
\begin{center}
\includegraphics[width=0.99\linewidth]{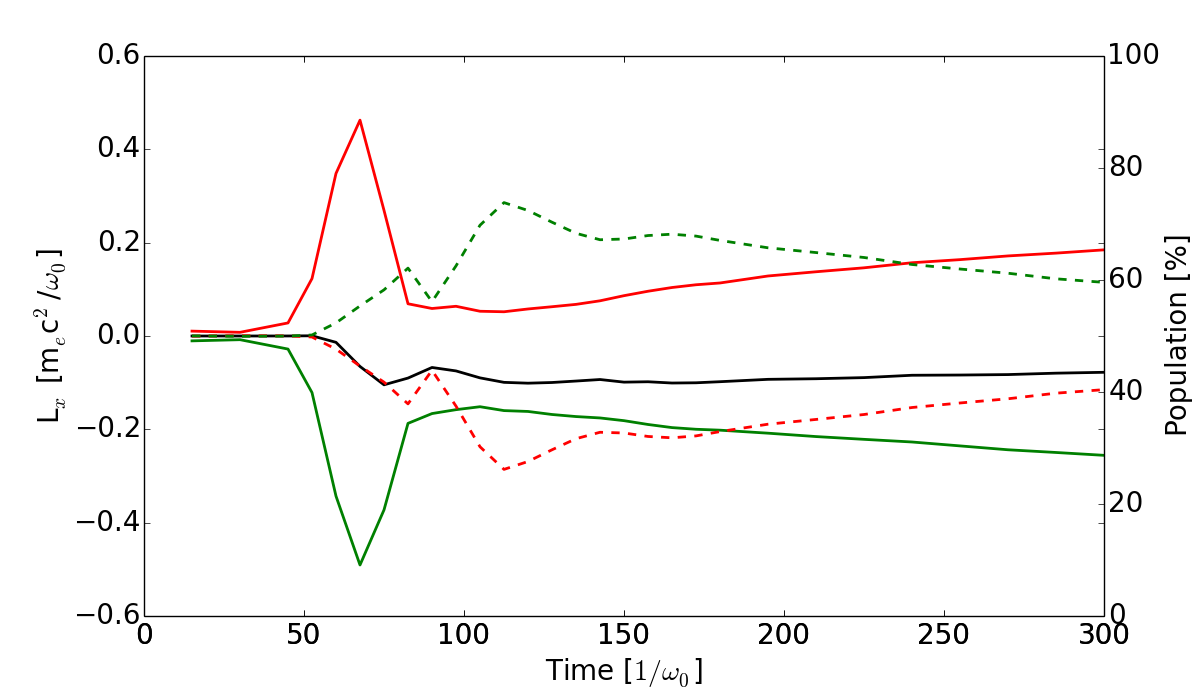}
\caption{Longitudinal orbital angular momentum averaged over all the electron population 
(black solid curve), over electrons with positive $\langle L_x \rangle$ (red solid curve) and 
negative $\langle L_x \rangle$ (green solid curve). Fraction of electron population 
with positive (red dashed curve) and negative (green dashed curve) $\langle L_x \rangle$. }
\label{fig:Lx_rad}
\end{center}
\end{figure}

From PIC simulations, we evaluate the averaged longitudinal orbital angular momentum 
gained by electron in the laser beam:
\begin{equation}
\langle L_x \rangle  = \frac{1}{N_b}\sum_{i=1, N_b} w_i[(y_i-y_0) \times p_{z,i} - (z_i-z_0) \times p_{y,i}]
\end{equation}
with the total number of electrons $N_b$, laser beam axis (y$_0$, z$_0$), transverse coordinates 
($y_i$,$z_i$) and transverse momenta (p$_{y,i}$, p$_{z,i}$) of the macro-particle $i$ defined with the weight $w_i$. Protons do not gain any significant angular momentum
 due to their mass 1836 times heavier than the electron's one. Solid curves in Fig. \ref{fig:Lx_rad} display the temporal evolution of $\langle L_x \rangle$: the red and green curves, 
distinguishing electrons rotating in opposite directions, show a similar temporal behaviour but of an 
opposite sign. As the laser interacts with the plasma bulk, from $\omega_0t$ = 50 to  
$\omega_0t$ = 100 ($1/\omega_0 = 0.4248$ fs), 
the absolute value of $\langle L_x \rangle$ increases up to $0.5\ m_ec^2/\omega_0$, and 
then decreases to a value close to $0.15-0.2\ m_ec^2/\omega_0$. The black curve, displaying the 
total averaged electron angular momentum, exhibits a break around $\omega_0t \simeq 70$. $\langle L_x \rangle$ remains nonzero once the electron-laser interaction
ends: equal to -0.08 $m_ec^2$/$\omega_0$ at $\omega_0t \simeq 75$, it slightly increases up to 
-0.063 $m_ec^2$/$\omega_0$ at $\omega_0t \simeq 250$. 
The dashed curves in Fig. \ref{fig:Lx_rad}, displaying the relative number of electrons with
positive and negative angular momentum, show that the part of electrons with negative 
angular momentum continuously increases during their interaction with the laser. 
At the end of their interaction with the laser, more than 70\% of electrons have acquired 
a negative angular momentum. 

\begin{figure}[h]
\begin{center}
\includegraphics[width=0.99\linewidth]{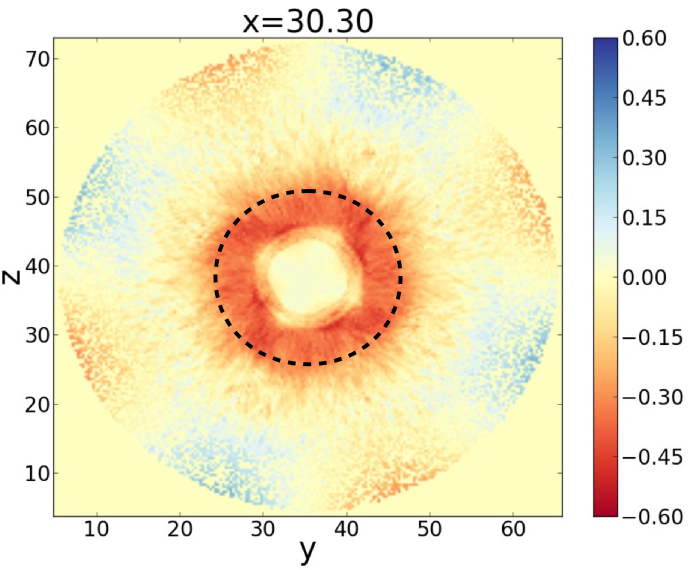}
\caption{Transversal cut of the electron $L_x$ distribution near x = 30 $c/\omega_0$ for 
$\omega_0$t = 120. The dashed black curve displays the field maximum amplitude contour.}
\label{fig:Lx_cutx}
\end{center}
\end{figure}

The transversal cut of the electron longitudinal OAM distribution, shown 
in Fig. \ref{fig:Lx_cutx}, displays a ring (a tube in 3D) containing the highest 
electron $\langle L_x \rangle$ amplitude with an external radius $r_1 \sim  15\ c/\omega_0$ 
and an internal one $r_0 \sim c/\omega_0$. These rotating electrons, characterized
by $\langle L_x \rangle$ slightly lower than -0.4 $m_ec^2/\omega_0$, form a solenoidal 
structure with an axial magnetic field inside (left panel in Fig.\ref{fig:field}). 
The central hole shows electrons $L_x$ close to zero values, in agreement with the laser intensity distribution. 

The left panel in Fig. \ref{fig:field}  displays an isocontour of the longitudinal magnetic field 
$B_x$ once 
the laser has left the plasma area. The quasi-static $B_x$ field presents a cylindrical shape: 
its axial length is equal to the plasma length ($\sim 2.5\ \mu$m) and its radius 
is limited by the contour where the laser intensity is maximal ($r=12\ c/\omega_0$). The cylinder axis coincides with 
the laser one. The $B_x$ maximal amplitude reaches 0.004 $m_e \omega_0/e$ (53 T).

We now compare this value with the $B_x$ field generated by electrons localized in 
a hollow cylinder with a length $l$, an internal radius $r_0$ and an external radius $r_1$. 
The electron density is set to $n_e$ and the electron longitudinal orbital angular
momentum is equal to $l_x$. We consider the laser unit system, where lengths are expressed
in $c/\omega_0$=127.3 nm, density is expressed in $n_c=1.74 \times 10^{19} $cm$^{-3}$ (the plasma critical density) and magnetic field is expressed in $B_c = m_e \omega_0 / e = 13382$ Tesla. 
The derivation of the Biot-Savart law \cite{jackson}, presented in the Supplementary
Materials, results in 
\begin{equation}
B_x \simeq - 0.5 n_e l_x \Big[\frac{l}{r_0}-\frac{l}{r_1}\Big]
\label{eq:Bx}
\end{equation}
Considering the values computed in PIC simulations: $\langle L_x \rangle =-0.4\ m_ec^2/\omega_0$, 
$n_e = 0.01$ $n_c$, $l$ = 19 $c/\omega_0$, $r_0$=6 $c/\omega_0$ and $r_1$=15 $c/\omega_0$, 
we obtain $B_x$ = 0.0038 $B_c$ which perfectly agrees with the $B_x$ value 
computed from PIC simulations and equal to 0.004 $B_c$. 

\begin{figure}[h]
\begin{center}
\includegraphics[width=0.99\linewidth]{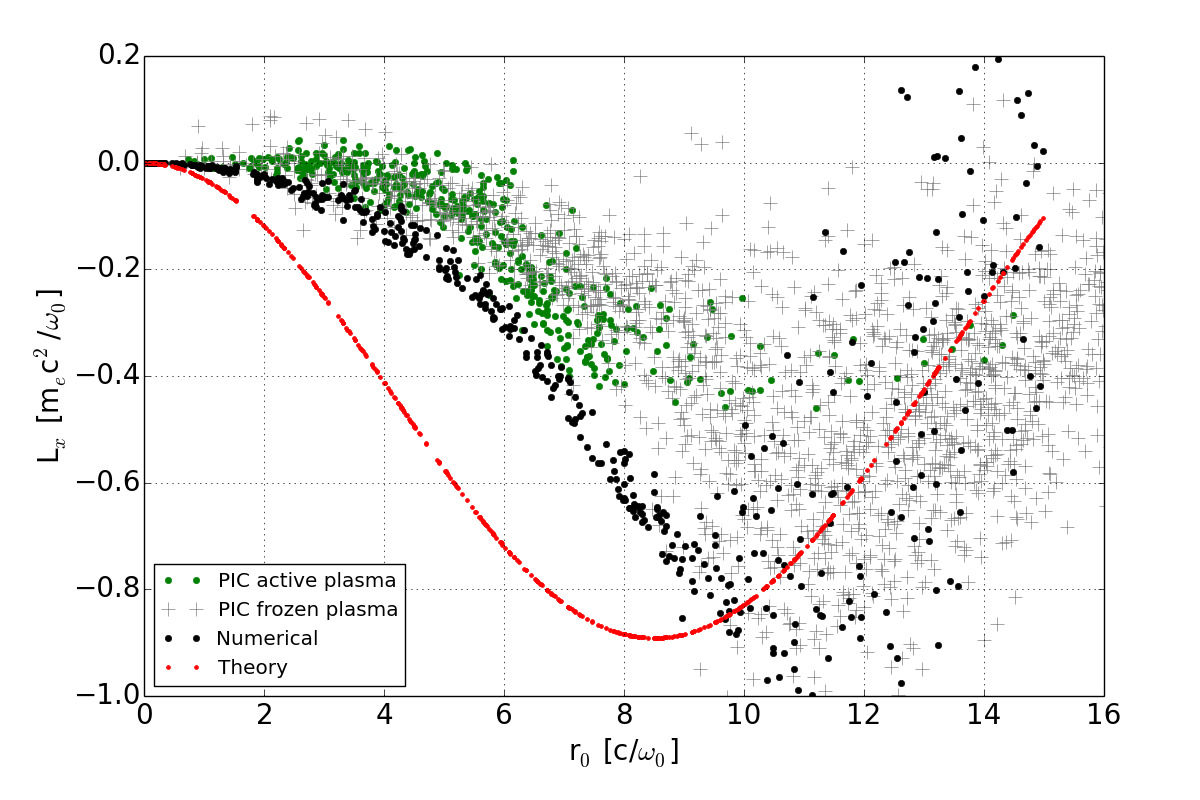}
\caption{Longitudinal orbital angular momentum ($L_x$) computed at 
$\omega_0$t = 97 versus the initial electron radial position with (green circle) 
and without (gray markers) active plasma motion. The red dotted curve and the black circles 
display the $L_x$ values computed, analytically and 
numerically, respectively, by using the perturbation theory.}
\label{fig:ptcl_Lxr0}
\end{center}
\end{figure}

To understand how electrons gain their angular velocity, we have implemented in the OCEAN code 
a particle tracking module. This diagnostic consists in adding some arbitrarly chosen electrons 
that experience the electric and magnetic fields but do not produce 
the self-consistent fields, automatically computed in PIC simulations. Figure \ref{fig:ptcl_Lxr0} 
displays the longitudinal OAM acquired by these ``test electrons'' at the end of their interaction with the laser beam as a 
function of their initial radial coordinate ($r_0$), when the plasma motion is active (green circle) so that 
plasma self-consistent fields influence the laser-electron dynamics and when the plasma motion is ``frozen''
(gray markers) where these ``test electrons'' only experience the laser electromagnetic fields. 

The green circles show that $L_x$ values lower than  $-0.4\ m_ec^2/\omega_0$ are obtained for electrons 
initially localized in the zone where the laser intensity is maximum, i.e. $r = 7 - 12 \ c/\omega_0$. 
Moreover, the $L_x$ spatial distribution follows the laser intensity spatial shape.
Same characteristics are observed for the $L_x$ radial distribution when the
plasma motion is frozen (gray markers), except that the electron angular momenta reach higher 
absolute values. 
This demonstrates that the OAM transfer from laser beam to electrons is a pure optical process
for which the efficiency is increased as the plasma self-consistent fields are turned off. When the plasma motion is active, a longitudinal charge separating field 
is generated on the rear plasma surface due to the ion inertia. This electrostatic field keeps the
electrons confined in the plasma zone reducing their interaction time with the laser, and then their 
angular velocity gain.

Because these PIC simulations clearly highlight that the laser to electrons OAM transfer results from
optical process only, we develop a reduced numerical model describing the laser beam induced electron dynamics. 
It solves the relativistic motion 
equations of an electron irradiated by the electric ($E_r$, $E_\theta$, $E_x$) and 
magnetic fields ($B_r$, $B_\theta$, $B_x$).
\begin{eqnarray}
\dot{p_r} & = & - E_r - v_{\theta}B_x + v_{x}B_{\theta}+ p_{\theta} \dot{\theta}  \label{eq:vr}\\
\dot{p_{\theta}}& =&  -  E_{\theta} - v_x B_r + v_rB_x - p_r \dot{\theta}         \label{eq:vtheta}\\
\dot{p_{x}}    & =&  - E_x - v_r B_{\theta} + v_{\theta} B_r \label{eq:vx}
\end{eqnarray}
where $p_r =\gamma v_r= \gamma \dot{r}$, $p_\theta =\gamma v_\theta= \gamma r\dot{\theta}$ and 
$p_x =\gamma v_x=\dot{x}$ are the radial, azimuthal and longitudinal electron momenta, respectively and $\gamma = \sqrt{1+p_r^2+p_\theta^2+p_x^2}$ is the Lorentz factor. 
As detailed in the Supplementaty Materials, a radially polarized OAM laser beam is not an eigenmode of 
the Maxwell's equations: it contains a small admixture of the azimuthal component out of the focal point. 
The electric and magnetic field components are then approximated with :
\begin{align}
E_r&= a_0 f(r) g(t)\cos(\phi) & B_r & =  -l a_0 \alpha f(r)g(t)\cos(\phi) \label{eq:EBr}\\
E_\theta& =l a_0 \alpha f(r) g(t)\cos(\phi) & B_\theta & =  a_0 f(r)g(t)\cos(\phi) 
\label{eq:EBtheta}\\
E_x & =  a_0 \frac{f(r)}{r}g(t)\sin(\phi) & B_x & =  - l a_0 \frac{f(r)}{r}g(t)\cos(\phi) 
\label{eq:EBx}
\end{align}

where $\phi = \omega_0 t -l\theta -x$ is the phase and $\alpha <<1$ is the parameter accounting for the contribution
of the additional azimuthal component generated out of the focal point.
First, we assume a low laser intensity ($a_0 << 1$) which allows a perturbative expansion of
Eqs.(\ref{eq:vr}, \ref{eq:vtheta}, \ref{eq:vx}) in powers of $a_0$ 
(details are given in Supplementary Materials). Integrated over the pulse duration, the first order of electron 
OAM is zero, so that it does not contribute to the final electron OAM. The second order term of $L_x$ solves :
\begin{equation}
\dot{L_x}^{(2)}= r^{(1)}\dot{p_\theta}^{(1)}-r_0p_x^{(1)}B_r + r_0p_r^{(1)}B_x - 
r_0 \frac{d E_\theta}{d\vec{r}}(\vec{r}^{(1)}-\vec{r_0})
\end{equation}
In fact, only the $-r_0 \frac{d E_\theta}{dr}(r^{(1)}-r_0)$ and $-r_0 \frac{d E_\theta}{dx}(x^{(1)}-x_0)$ terms finally
contribute to :
\begin{equation}
L_x^{(2)}(\tau) = -\alpha l a_0^2 f^2(r_0) \frac{3\tau}{8} \bigg[1-\bigg(\frac{r_0}{w_0}\bigg)^2  \bigg]
\label{eq:Lx}
\end{equation}

\begin{figure}[h]
\begin{center}
\includegraphics[width=0.99\linewidth]{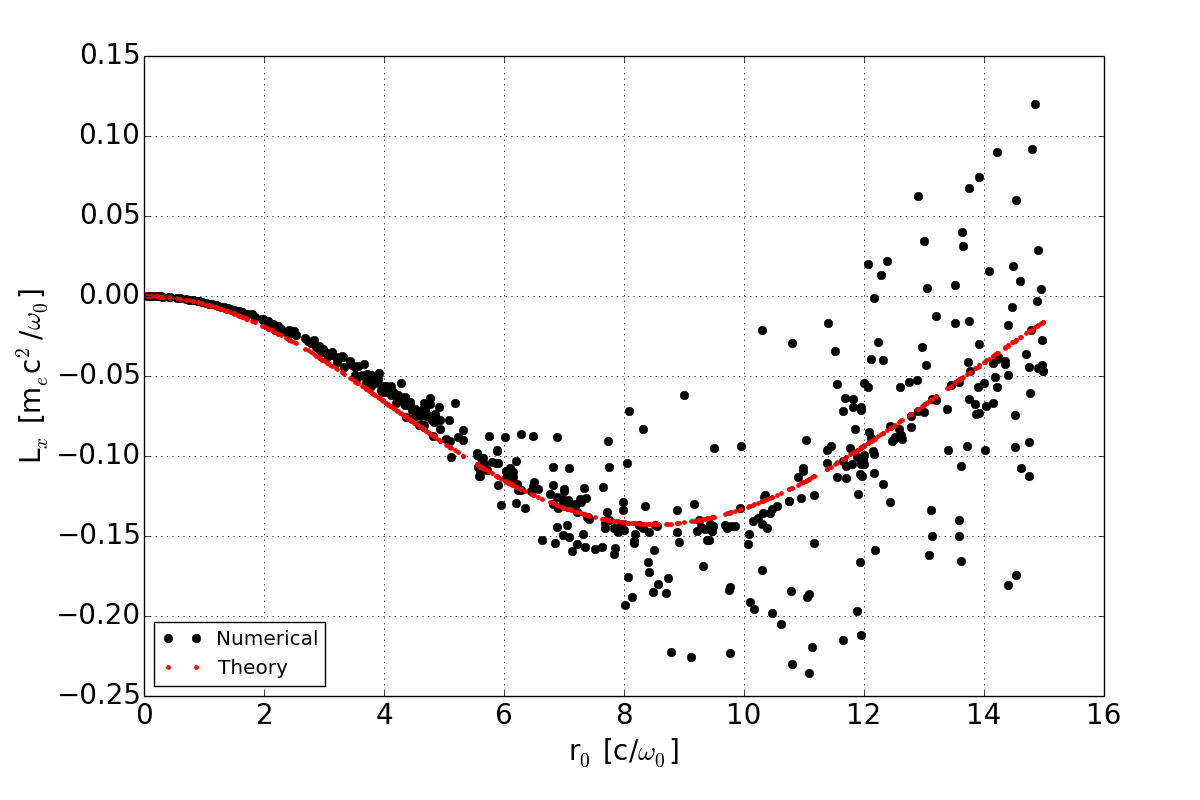}
\caption{Longitudinal orbital angular momentum of electrons versus their initial radial position computed 
from the numerical integration of motion equations (black circle) and the theoretical prediction according to
Eq. \ref{eq:Lx} (red curve) for $a_0 = 0.4$ and $\alpha$ = 0.1. 
The other laser parameters are identical to the PIC simulations : $w_0 = 2\ \mu$m, $\tau = 6$ T, $l=1$.}
\label{fig:ptcl_Lx_python_a0_0_4}
\end{center}
\end{figure}

Figure \ref{fig:ptcl_Lx_python_a0_0_4} exhibits a perfect agreement between
the electron $L_x$ radial distribution computed from the
numerical integration of Eqs.(\ref{eq:vr}, \ref{eq:vtheta}, \ref{eq:vx}) and from the analytical
expression (Eq.(\ref{eq:Lx})) for $a_0 = 0.4$. 
As already observed in the PIC simulations [Fig. \ref{fig:ptcl_Lxr0}],
even at low intensity, most of the electrons acquire a negative longitudinal OAM. This analysis
evidences that the laser to electron OAM transfer originates from the radial and longitudinal electron drift 
in the laser field. For $r_0=w_0/\sqrt{2}$ (where the intensity is maximum), the electrons acquire the maximum angular 
rotation, and the theoretical expression of $L_x^{(2)}$ shows that the electron longitudinal motion
contributes for 2/3 and the radial one for 1/3 to the final angular momentum value.

For higher laser intensity, $a_0 = 1$, the black circles and the red curve in Fig. \ref{fig:ptcl_Lxr0} 
display the $L_x^{(2)}$ values computed by numerically integrating Eqs. (\ref{eq:vr},\ref{eq:vtheta},\ref{eq:vx})
and with the theoretical formula (Eq. (\ref{eq:Lx})), respectively. The analytical expression deviates
from the numerical integration of motion equations, but demonstrates a qualitative similar behaviour. We 
observe negative longitudinal OAM gained by electrons, and the correct order of magnitude for the $L_x$ values.
A relatively good agreement between numerical results and PIC simulations is evident even if the azimuthal
component of the field is chosen constant along the laser propagation in the numerical model
[see Supplementary materials for more details]. In both cases, the laser transfers to electrons 
negative $L_x$ values, and the OAM transfer efficiency follows the laser intensity radial distribution.

\begin{figure}[h]
\begin{center}
\includegraphics[width=0.99\linewidth]{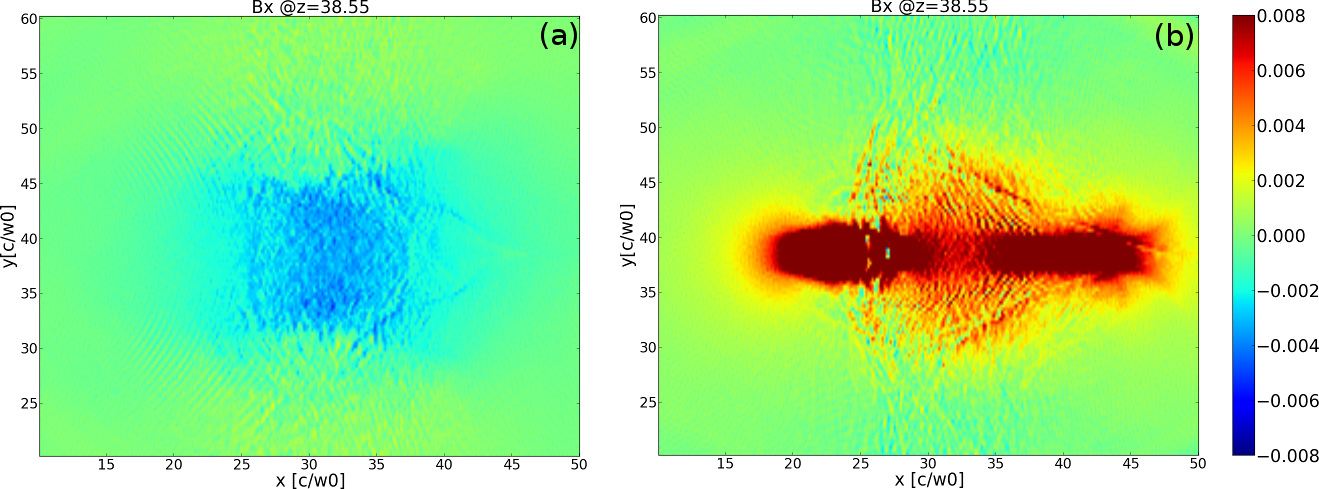}
\caption{Longitudinal cut of $B_x$ computed from PIC simulations for $l=-1$ and $\tau = 6$ T (a) and
$l=1$ and $\tau = 12$ T (b). The others laser parameters are kept unchanged.}
\label{fig:Bx_12T}
\end{center}
\end{figure}

The direction of the electron OAM changes by inverting the sign of the OAM laser beam [see Supplementary materials]. 
The generated solenoidal magnetic field is then directed to the opposite 
direction [see Fig.\ref{fig:Bx_12T}(a)]. 

As suggested by Eq. (\ref{eq:Lx}), the electron OAM amplitude is higher as the pulse duration is increased
[see Supplementary materials]. Figure \ref{fig:Bx_12T}(b), displaying the longitudinal cut of $B_x$ computed 
with a pulse duration equal to 12 optical cycles, shows that its amplitude is two times higher than for a pulse 
duration equal to 6 optical cycles. The longitudinal extension of the magnetic
field is larger than for the shorter pulse duration, because the electrons have acquired 
higher longitudinal momenta resulting in a larger plasma expansion.

In conclusion, we demonstrated that a tightly focused radially polarized laser beam is able to transfer its 
Orbital Angular Momentum to 
electrons without any dissipative effects like collision or ionization. Only an optical process is 
responsible for the electron rotation gain, which originates from the radial and longitudinal electron 
drift in the laser field. Both radial and azimuthal components of the laser fields are needed for the
 laser to electron OAM transfer. By using 3D Particle In Cell simulations, we have shown that the rotated electrons 
produce a quasi-static magnetic field $\sim$  50 T over more than 80 fs after the end of the laser-plasma
interaction. This magnetic field is homogeneous over spatial dimensions (a 2.5 $\mu$m length and a 1.5 $\mu$m as transverse size) defined by the plasma length and the laser transverse size. An accurate choice of
laser parameters, such as focal beam waist, laser intensity, laser OAM and/or pulse duration, make then possible a 
control of the quasi-static magnetic field production. 
A simple analytical model provides a quantitative estimation of the electron longitudinal orbital angular momentum
 values and the solenoidal magnetic field up to $a_0 = 1$. This work opens new ways to optically generate 
quasi-static magnetic field from the laser plasma interaction.

This work was granted access to HPC resources of TGCC under the allocation A0010506129 made by GENCI.

%\bibliographystyle{unsrt}
%\bibliography{../../Bibliography}

\end{document}